\pdfoutput=1

\documentclass[11pt]{article}

\usepackage[preprint]{acl}

\usepackage{times}
\usepackage{latexsym}

\usepackage[T1]{fontenc}

\usepackage[utf8]{inputenc}

\usepackage{microtype}

\usepackage{inconsolata}

\usepackage{graphicx}

\usepackage{booktabs}
\usepackage{amssymb}
\usepackage[normalem]{ulem}
\useunder{\uline}{\ul}{}

\title{Eliciting Instruction-tuned Code Language Models' Capabilities to Utilize Auxiliary Function for Code Generation}

\author{Seonghyeon Lee\textsuperscript{1}, Suyeon Kim\textsuperscript{1}, Joonwon Jang\textsuperscript{2}, Heejae Chon\textsuperscript{3}, Dongha Lee\textsuperscript{3}, Hwanjo Yu\textsuperscript{1}\thanks{\ \ Corresponding author} \\
        Department of Computer Science and Engineering, POSTECH, Pohang, South Korea\textsuperscript{1} \\
        Department of Artificial Intelligence, POSTECH, Pohang, South Korea\textsuperscript{2} \\
        Department of Artificial Intelligence, Yonsei University, Seoul, South Korea\textsuperscript{3} \\
        \texttt{sh0416@postech.ac.kr}}

\begin{document}
\maketitle
\begin{abstract}
We study the code generation behavior of instruction-tuned models built on top of code pre-trained language models when they could access an auxiliary function to implement a function.
We design several ways to provide auxiliary functions to the models by adding them to the query or providing a response prefix to incorporate the ability to utilize auxiliary functions with the instruction-following capability.
Our experimental results show the effectiveness of combining the base models' auxiliary function utilization ability with the instruction following ability.
In particular, the performance of adopting our approaches with the open-sourced language models surpasses that of the recent powerful proprietary language models, i.e., \texttt{gpt-4o}.
\end{abstract}

\section{Introduction}

Generating codes based on natural language requirements, i.e., code generation, becomes an appealing application for natural language processing community due to the recent advance of code pre-trained language models~\citep{singh-etal-2023-codefusion,zhou-etal-2023-codebertscore,wang-etal-2023-codet5,zhang-etal-2023-repocoder}.
Pre-training on large-scale code corpora enables a language model to implement correct functions based on their requirements written in the docstrings.
Also, tuning a code pre-trained language model to follow instructions has been released due to the effectiveness of instruction-tuned language models on natural language tasks~\citep{luo2024wizardcoder,wei2023magicoder,song2024alchemistcoder,lei2024autocoder}.\footnote{From now on, we call an instruction-tuned code pre-trained model an instruction-tuned model for brevity.}
These instruction-tuned models boost up the code generation ability.

In code generation tasks, leveraging an auxiliary function reduces the implementation difficulty of a target function compared to that of implementing them from scratch.
The auxiliary function is a function that helps implement a target function by inspiring novel mechanism or handling complicated subroutines for the target function through function calls~\citep{lee2024exploring}.
Therefore, properly utilizing the given auxiliary function becomes important for the instruction-tuned models.

However, limited research has been conducted on providing auxiliary functions to make the instruction-tuned models utilize the auxiliary functions effectively.
\citet{lee2024exploring} initially included the auxiliary function in the prompt, but it showed inferior results compared to just prompting the corresponding base pre-trained models.
Also, the instruction-tuned models' ability to incorporate the code content with their natural language text has not been fully explored, except that the model providers showcase some qualitative examples in their appendix~\citep{rozière2024code}.

In this work, we comprehensively explore the instruction-tuned models' code generation behavior when they can access an auxiliary function.
To do this, we design several prompts that are likely to elicit the ability to utilize auxiliary functions by leveraging the query-response structure employed in the instruction-tuned models.
To be specific, we provide detailed information about the auxiliary function in the query and provide an incomplete codeblock to the prefix in the response to complete the remaining response.
Then, we evaluate their effectiveness across several competitive instruction-tuned models.
Our evaluation results show that our proposed prompts perform efficaciously on the instruction-tuned models compared to the corresponding base models, and even surpass \texttt{gpt-4o}, which is purportedly known as the most powerful language model.
Finally, we perform an in-depth analysis to demonstrate that incorporating auxiliary function utilization ability already encoded in their base model with instruction-following capability through the response prefix is the main cause of the superior performance.

\begin{figure}[t]
  \includegraphics[width=\columnwidth]{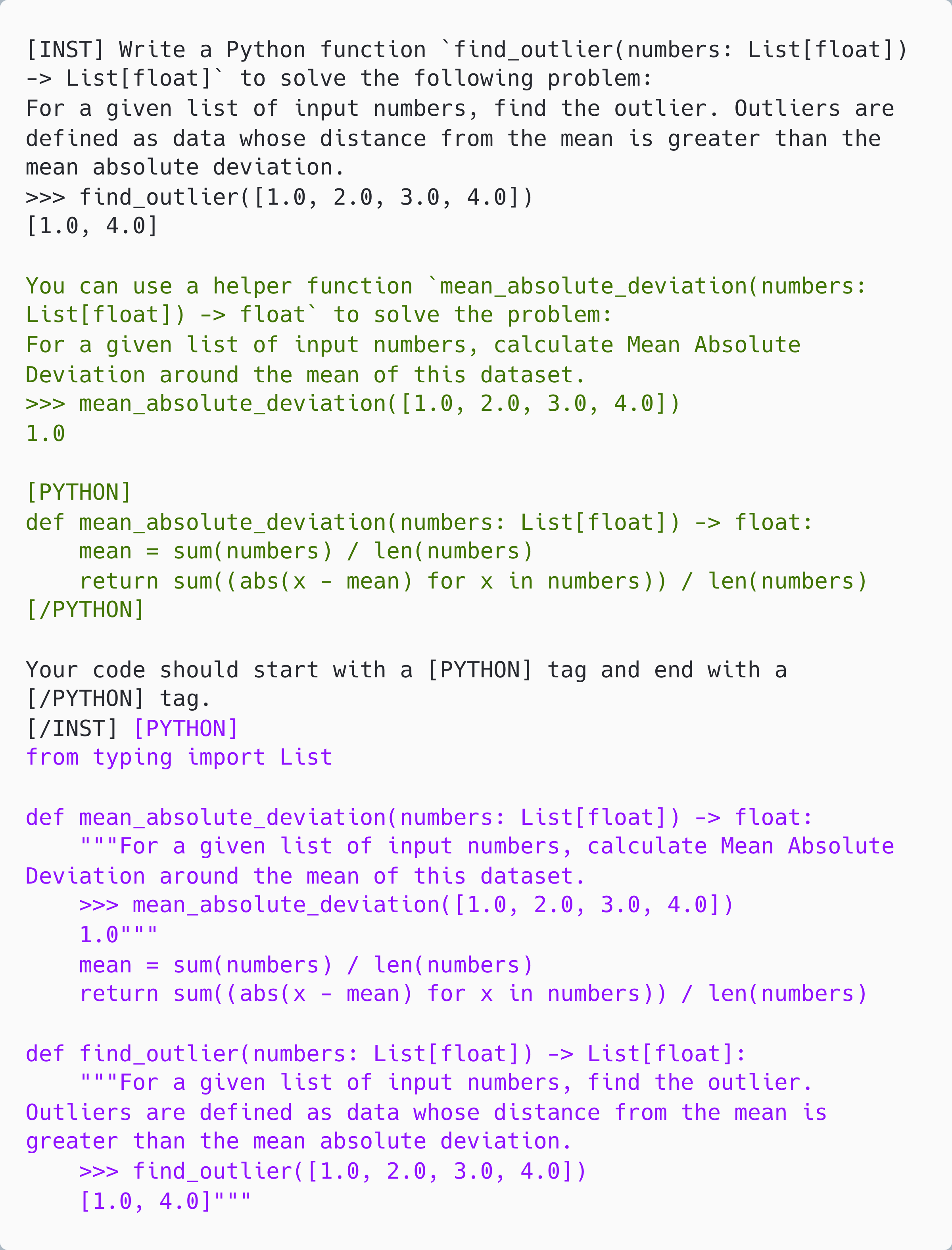}
  \caption{An example of our proposed prompts compatible with \texttt{CodeLlama-Inst} format. The green and purple part indicate inserting information about auxiliary function into the query and response, respectively.}
  \label{fig:prompt-example}
\end{figure}

\section{Methods}
We design several prompts for pre-trained models and instruction-tuned models to show their ability to generate code with and without auxiliary functions.
To do this, we build up two different approaches to naturally fuse the information inside the auxiliary functions into the instruction format used during instruction tuning.

\subsection{Preliminary}
We briefly explain how the existing work uses instruction-tuned models to implement codes.
In general, the instruction-tuned models comply with a query-response format.
Within this format, the models are trained to respond to that query.
Therefore, we simply write a query with the given requirements to induce them to implement the code.
The query consists of (1) the objective statement, (2) a description of a function, and (3) a formatting guideline.
The objective statement commands them to implement a function, and a description of the function is followed to explain their functionality with some examples.
Finally, the output formatting guideline is provided to easily parse the codeblock from their response.
The constructed query is inserted into the pre-defined format for the model prompts.
Then, the models generate a response that contains a codeblock with the implementation.

\subsection{Incorporating the function utilization ability with instruction-following ability}
On top of the existing approach, we propose simple and effective ways to enhance the instruction-tuned models' ability to utilize other functions.

\paragraph{Approach 1. Inserting auxiliary function information in the query-side}
The first approach (Figure~\ref{fig:prompt-example}, Green) is to insert the information about auxiliary function to the query.
Assuming that the instruction-tuned models understand the codes in their query, we can expect that information about the auxiliary function, e.g., declaration, docstring, and their implementation, provided in the query would be comprehended by the models and leveraged when generating their response.

\paragraph{Approach 2. Inserting auxiliary function definition in the response-side}
Another approach (Figure~\ref{fig:prompt-example}, Purple) is to attach an informative starting text to the response to guide the models to naturally complete the remaining response.
One similar approach that has been conducted in general tasks is to elicit language models' reasoning capability by attaching a starting text, e.g., let's think step by step, to the response~\citep{NEURIPS2022_8bb0d291}.
Motivated by this work, we attach the incomplete codeblock that models should generate within their response.
Specifically, we open a codeblock with an appropriate tag and function declaration requested to be implemented in the query with proper import statements.
Using this approach also guarantees that the codeblock is properly opened, so the task for the models becomes easier as they just properly finalize the codeblock.
In addition, we insert an auxiliary function into the codeblock for the models to leverage the auxiliary function during the generation.

\section{Experiments}
We explore the effectiveness of our proposed strategy with recent competitive instruction-tuned models and analyze their behavior from various angles.

\begin{table*}[t]
\centering
\resizebox{\textwidth}{!}{%
\begin{tabular}{llll}
\toprule
Base model & No Aux & Aux & Instruction-tuned variants \\ \midrule
\texttt{CodeLlama-7b}~\citep{rozière2024code} & 0.1973 & 0.5284 & \texttt{CodeLlama-Inst-7b}, \texttt{MagicoderS-CL-7b} \\
\texttt{Deepseek-coder-6.7b}~\citep{guo2024deepseekcoder} & 0.2688 & 0.6741 & \texttt{Deepseek-coder-Inst-6.7b}, \texttt{MagicoderS-DS-6.7b} \\
\texttt{Deepseek-coder-7b}~\citep{guo2024deepseekcoder} & 0.3066 & 0.6152 & \texttt{Deepseek-coder-Inst-7b} \\
\texttt{CodeGemma-7b}~\citep{codegemma} & 0.2623 & 0.6043 & \texttt{CodeGemma-Inst-7b}, \texttt{CodeGemma-Inst-1.1-7b} \\
\texttt{Llama3-8b}~\citep{llama3modelcard} & 0.1828 & 0.4914 & \texttt{Llama3-Inst-8b} \\ \midrule
\texttt{Starcoder2-15b}~\citep{lozhkov2024starcoder} & 0.3066 & 0.6682 & \texttt{Starcoder2-Inst-15b} \\
\texttt{CodeLlama-34b}~\citep{rozière2024code} & 0.2709	& 0.6411 & \texttt{CodeLlama-Inst-34b} \\
\texttt{Deepseek-coder-33b}~\citep{guo2024deepseekcoder} & 0.3599 & 0.7248 & \texttt{Deepseek-coder-Inst-33b} \\ \bottomrule
\end{tabular}%
}
\caption{Base models and their instruction-tuned variants. We measure the pass@1 score of the base models on Humanextension to correctly identify whether our approaches can acquire better performance compared to that of prompting the base model.}
\label{tab:models}
\end{table*}

\begin{table*}[t]
\centering
\resizebox{\textwidth}{!}{%
\begin{tabular}{@{}lcccccc@{}}
\toprule
Instruction-tuned model & \multicolumn{2}{c}{w/o auxiliary function} & \multicolumn{4}{c}{w/ auxiliary function} \\ \midrule
1. Insert auxiliary function info to query & & & \checkmark & \checkmark & & \checkmark \\
2.1. Insert incomplete codeblock to response & & \checkmark & & \checkmark & \checkmark & \checkmark \\
2.2. Insert auxiliary function to incomplete codeblock & & & & & \checkmark & \checkmark \\ \midrule
\texttt{CodeLlama-Inst-7b}~\citep{rozière2024code} & 0.2907 & 0.2825 & 0.4844 & {\ul 0.5503} & {\ul \textbf{0.5583}} & {\ul 0.5477} \\
\texttt{MagicoderS-CL-7b}~\citep{wei2023magicoder} & 0.3977 & 0.4848 & 0.4656 & {\ul 0.6440} & {\ul \textbf{0.6550}} & {\ul 0.6437} \\
\texttt{Deepseek-coder-Inst-6.7b}~\citep{guo2024deepseekcoder} & 0.3990 & 0.5497 & 0.6613 & 0.6623 & {\ul \textbf{0.6894}} & {\ul 0.6828} \\
\texttt{MagicoderS-DS-6.7b}~\citep{wei2023magicoder} & 0.4934 & 0.5507 & 0.6325 & {\ul 0.7050} & {\ul 0.6828} & {\ul \textbf{0.7265}} \\
\texttt{Deepseek-coder-Inst-7b}~\citep{guo2024deepseekcoder} & 0.5348 & 0.6079 & {\ul 0.6414} & {\ul 0.6970} & {\ul 0.7166} & {\ul \textbf{0.7348}} \\
\texttt{CodeGemma-Inst-7b}~\citep{codegemma} & 0.3086 & 0.3881 & 0.5324 & {\ul 0.6083} & {\ul \textbf{0.6228}} & {\ul 0.6060} \\
\texttt{CodeGemma-Inst-1.1-7b}~\citep{codegemma} & 0.3354 & 0.4626 & 0.4563 & {\ul 0.5970} & {\ul \textbf{0.6219}} & {\ul 0.6182} \\
\texttt{Llama3-Inst-8b}~\citep{llama3modelcard} & 0.3632 & 0.3950 & {\ul 0.5801} & {\ul \textbf{0.5868}} & {\ul 0.4970} & {\ul 0.5772} \\ \midrule
\texttt{Starcoder2-Inst-15b}~\citep{lozhkov2024starcoder} & 0.4182 & 0.5116 & 0.6325 & {\ul \textbf{0.7079}} & {\ul 0.6834} & {\ul 0.6934} \\ 
\texttt{CodeLlama-34b-Inst~\citep{rozière2024code}} & 0.3599 & 0.3550 & 0.6219 & {\ul \textbf{0.6421}} & 0.6146 & 0.6364 \\
\texttt{Deepseek-coder-33b-Inst}~\citep{guo2024deepseekcoder} & 0.4904 & 0.5957 & 0.6546 & {\ul 0.7503} & {\ul 0.7510} & {\ul \textbf{0.7639}} \\ \midrule
\texttt{gpt-3.5-turbo-0125}~\citep{achiam2023gpt} & 0.4868 & & 0.5901 & & & \\ 
\texttt{gpt-4o-2024-05-13}~\citep{achiam2023gpt} & 0.6358 & & 0.6987 & & & \\ \bottomrule
\end{tabular}%
}
\caption{Humanextension pass@1 score for instruction-tuned models with the proposed prompts. We mark bold on the most effective score in each model and underline the scores that surpass their base models. For \texttt{gpt} models, we could not report some scores as a user is not technically allowed to add a prefix to the response side.}
\label{tab:main-table}
\end{table*}

\subsection{Experimental setup}

\paragraph{Basic setup}
We list the recent competitive instruction-tuned models and their corresponding base models in Table~\ref{tab:models}.
We adopt the Humanextension benchmark consist of 151 relevant function pairs specially designed for measuring the language models' ability to utilize other functions~\citep{lee2024exploring}.
We follow the widely used decoding strategy for generating code: 0.2 for temperature and 0.95 for top p, and generate at most 512 tokens per prompt~\citep{bigcode-evaluation-harness}.
We evaluate the generated implementations using functional correctness by measuring the proportion of correct implementations that pass whole test cases among 20 generations, which is known as pass@1 score.

\paragraph{Measuring the base models' auxiliary function utilization capability}
We measure the pre-trained base models' ability to utilize other functions using the Humanextension benchmark with the prompt where the detailed prompt for the base models can be found in the Appendix.
In doing so, we disentangle the strength of the instruction-tuned models from our experimental results by considering the improvement already observed in the base models. 
We compare this score to evaluate whether the proposed approaches with the instruction-tuned models could surpass this simple baseline.

\subsection{Results}

\paragraph{Overall results}
We demonstrate that our approaches successfully elicit the instruction-tuned models' ability to utilize auxiliary functions.
We report the pass@1 score with various prompting approaches in Table~\ref{tab:main-table}.
First, both adding information about auxiliary functions in the query and appending a codeblock with an auxiliary function definition in the response show a clear improvement compared to their closest counterparts.
Comparing the first and third columns or the second and fourth columns demonstrates the effectiveness of providing information about auxiliary functions to the query.
Also, comparing the second and fifth columns verifies the effectiveness of attaching the auxiliary function definition as a prefix for their response.
Furthermore, our proposed prompts also work effectively in the bigger sized model such as \texttt{Deepseek-coder-33b-Inst}.
We additionally report the performance of the powerful proprietary instruction-tuned models such as \texttt{gpt-3.5-turbo} and \texttt{gpt-4o} and verify that the open-sourced models can easily surpass them with our prompting approach.

\paragraph{Model analysis}
When we look at each model, \texttt{MagicoderS-CL-7b} shows superior performance compared to \texttt{CodeLlama-Inst-7b}, representing the importance of diverse instruction-tuned datasets and this trend is also observed in \texttt{Deepseek-coder-6.7b}.
Assuming that \texttt{Deepseek-coder-Inst-6.7b} and \texttt{Deepseek-coder-Inst-7b} are trained on the same instruction-tuned dataset, enhancing the base models is also a plausible direction to improve their code generation capability with auxiliary functions.
Comparing \texttt{CodeGemma-Inst-7b} and \texttt{CodeGemma-Inst-1.1-7b}, the pass@1 score is improved when they implement the given problem without an auxiliary function (from 0.3881 to 0.4626), but this improvement is not transferred when they access the auxiliary function (from 0.6228 to 0.6219).
\texttt{Llama3-Inst-8b} shows different patterns compared to other models in that their score mostly increases when the auxiliary function is in the query.
We speculate the reason for this results as \texttt{Llama3-8b} is pre-trained on the mixture of code and text corpora while other models focus only on code corpora.

\paragraph{Comparison with the base models}
We further examine the performance by comparing that of their corresponding base models.
We underscore the performance that surpasses the score that could be acquired by simply prompting the base models.
The scores in the last three columns mostly outperform their base models, demonstrating that applying both approaches at the same time successfully elicits the instruction-tuned models' auxiliary function utilization ability.
Based on our experimental results, the models generally surpass their base models when they gain knowledge about auxiliary functions with an incomplete codeblock in the response.


\begin{table}[]
\centering
\resizebox{\columnwidth}{!}{%
\begin{tabular}{@{}lcc@{}}
\toprule
Response prefix & \texttt{CodeLlama} & \texttt{CodeGemma} \\ \midrule
Add codeblock & 0.5503 & 0.6083 \\
Remove import statements & 0.5593 & 0.6159 \\
Remove docstring & 0.5060 & 0.5758 \\
Without codeblock & 0.4844 & 0.5324 \\ \bottomrule
\end{tabular}%
}
\caption{Pass@1 score with various response prefix content. \texttt{CodeLlama} shorts for \texttt{CodeLlama-Inst-7b} and \texttt{CodeGemma} shorts for \texttt{CodeGemma-Inst-7b}}
\label{tab:ablation-analysis}
\end{table}

\subsection{In-depth analysis for response codeblock}

We perform an in-depth analysis on the effectiveness of appending an incomplete codeblock to the response by dissecting the code-block into several components and observing the performance change as we remove them sequentially.
We report the performance in Table~\ref{tab:ablation-analysis}.
For \texttt{CodeLlama}, removing the docstring for the target function in the codeblock significantly drops the performance.
We conclude that \texttt{CodeLlama} understands the given requirements in the query with the docstring through the response codeblock.
On the other hands, \texttt{CodeGemma} preserves the performance after removing the docstring to some extent.
In this case, providing a function signature is much more crucial for \texttt{CodeGemma}.
From this results, we find that the instruction-tuned models focus on different code components and we further investigate this phenomena in future work.

\section{Conclusion}
In this work, we study the instruction-tuned models' behavior when they can access the auxiliary function. 
Through various prompting approaches, i.e., providing an auxiliary function in the query or response, we discover effective prompting approaches that enhance the probability of implementing correct codes using open-sourced models and surpass the recent powerful proprietary models, i.e., \texttt{gpt-4o}.
Our further investigation identifies that providing docstring or function signature to the response code-block is the major reason to boost performance.
We believe that incorporating the ability to utilize other functions with the instruction-following capability is indispensable for generating complex code, and our work becomes a cornerstone towards this research direction.

\section{Limitation}
There are a few limitations that have not been fully addressed in this work.
Due to the limited control of the proprietary models to the user, we could not report the score of \texttt{gpt-3.5-turbo} and \texttt{gpt-4o} when appending a prefix to the response which is verified as effective in the open-sourced models.
Also, whether the improvement made by the proposed approaches could be transferred through fine-tuning does not explore in this work, which will be our main future work.

\bibliography{anthology,custom}

\appendix
\section{Prompts for base models with auxiliary functions}
\label{sec:appendix}
We provide an illustrative prompt for evaluating the base models' ability to utilize auxiliary function are provided as follow.

\begin{verbatim}
from typing import List

def mean_absolute_deviation(numbers):
  """{auxiliary docstring}"""
  {auxiliary implementation}

def find_outliers(numbers):
  """{target docstring}"""
\end{verbatim}

We replace the content of auxiliary and target docstring and implementation for auxiliary function as placeholders for readability.
In this example, the base models implement \texttt{find\_outliers} and they could use \texttt{mean\_absolute\_deviation} as the function is given in the prompt.
\end{document}